\documentclass[11pt]{article}
\usepackage[top = 2 cm, bottom = 3 cm, left = 3 cm, right = 2 cm]{geometry}
\usepackage{authblk, cite, color, amssymb, amsmath}
\usepackage[hypertex, colorlinks = true, linkcolor = blue, citecolor = red]{hyperref}
\usepackage[dvipsnames]{xcolor}

\begin{document}

\title{\huge On the singular coordinate transformations of the Schwarzschild metric}

\author{\bf Merab Gogberashvili}
\affil{\small Javakhishvili Tbilisi State University, 3 Chavchavadze Avenue, Tbilisi 0179, Georgia \authorcr
Andronikashvili Institute of Physics, 6 Tamarashvili Street, Tbilisi 0177, Georgia}

\maketitle

\begin{abstract}
It is noted that the coordinate transformations usually used to demonstrate the continuity of geodesics at the Schwarzschild horizon are of class $C^0$, while the standard causality theory requires that the metric tensor to be at least $C^1$. Then this singular metric tensor leads to the appearance of the fictitious delta-like source in the Einstein equations, which prevents quantum particles to enter a black hole horizon.

\vskip 1mm

\end{abstract}
\vskip 5mm


Spacetime singularities are inevitable feature of many solutions of the Einstein equations \cite{Haw-Pen}. The opinion that they are the result of the high degree of symmetry, or are nonphysical in some way, contradicts to the famous singularity theorems \cite{San-Gar, Haw-Ell}. These theorems show geodesic incompleteness (currently the most acceptable criterion of existence of geometrical singularities) and assume that so called coordinate singularities (for which the curvature invariants are finite) can be avoided "repairing" geodesics by specific singular coordinate transformations. However, singularity theorems say little about the nature of the singularity, in some cases the curvature blows up when differentiability dropping below the class $C^2$.

Recall that a function is said to be of class $C^k$ if its derivatives, up to the order $k$, exist and are continuous. For example, the class $C^0$ consists of all continuous functions, while the class $C^1$ consists of all differentiable functions whose derivative exists and is of class $C^0$. For a class $C^0$ function, in general,
\begin{equation} \label{f'}
\int f'(x) dx \ne f(x)~,
\end{equation}
if the integration area contains a point of discontinuity of $f'(x)$, some kind of regularization procedure at this point is needed.

The $C^2$-differentiability assumption plays a key role in the singularity theorems. Indeed, the standard causality theory assumes that the metric is smooth. or is at least $C^2$, see \cite{Haw-Ell, Man-San, Clarke, Gar-San, Chru, Sor-Woo} for a reviews. Some authors assume that the components of the metric tensor can be of class $C^1$ (i.e. the second derivatives of the metric tensor, and thus the Riemann tensor, suffer a jump discontinuity), since there are many physically realistic systems of that type, such as the Oppenheimer-Snyder model of a collapsing star \cite{Opp-Sny} and general matched spacetimes, see e.g. \cite{Lich, Synge}. The issue of regularity in the singularity theorems is often ignored despite its mathematical and physical relevance \cite{Sano}.


We note that spacetime singularities should be associated with reaching the limits of the physical validity of general relativity, i.e. quantum effects can be expected to come in. Close to a singularity gravitational field becomes strong and some physical quantities diverge. Then, in our opinion \cite{Gog-Pan, Gog, Gog-Mod}, it is insufficient to explore this region only with the classical geodesic (Hamilton-Jacobi) equations,
\begin{equation} \label{geodesic}
g_{\mu\nu}p^\mu p^\nu - m^2 = 0~,
\end{equation}
where $p^\nu = m ~dx^\nu/ds$ denotes relativistic 4-momentum that contain the first derivatives of coordinate functions. Instead, one should use at least quasi-classical approximation and then obtain classical trajectories from the wavefunctions of quantum particles in geometrical-optical limit (eikonal approximation) \cite{Gold}. Einstein's gravity doesn't care about spin and close to a singularity particle trajectories can be described, for instance, with the solutions of the Klein-Gordon equation in a curve spacetime,
\begin{equation} \label{wave}
(\Box + m^2)\Phi = \left[\frac {1}{ \sqrt{-g}}\partial_\mu \left( \sqrt{-g}g^{\mu\nu}\partial_\nu \right) + m^2 \right]\Phi = 0~.
\end{equation}
For photons $m = 0$, while for the case of fermions one needs to obtain the equation of the type (\ref{wave}) from the first order Dirac's system.

It is known that the Klein-Gordon wave functions associated with the classical motion formally obey the relativistic Hamilton-Jacobi equation (\ref{geodesic}) written for the same system \cite{Motz}. Indeed, in the quasi-classical approximation the scalar wave function in (\ref{wave}) can be expressed in terms of an amplitude and phase, $\Phi = A \exp(iS)$, where $S$ is the Hamilton principal function, which usually is used in the definition of the classical momentum, $p^\nu \sim \partial^\nu S $. Then the Klein-Gordon equation (\ref{wave}) reduces to the system of equations:
\begin{eqnarray} \label{KG-system}
A \Box S + 2 \partial_\nu S \partial^\nu A = 0~, \nonumber \\
\Box A - A \partial_\nu S \partial^\nu S + m^2 A = 0~.
\end{eqnarray}
The approximations needed to reduce this system to the geodesic equation (\ref{geodesic}) are: (i) Week gravitational field and short wavelength, $\Box S \to 0$; (ii) Negligible variations of the wave amplitude, $\partial^\nu A \to 0$ \cite{Gog-Pan, Gog, Gog-Mod}. From the condition (i) it follows that the eikonal phase (Hamilton's principal function) can be written as $S \sim p_\nu x^\nu$, where $p_\nu$ obeys (\ref{geodesic}). Close to a singularity the approximations (i) and (ii) are not valid and to explore this region one needs to consider the Klein-Gordon equation (\ref{wave}), not the geodesic equations (\ref{geodesic}). Unlike (\ref{geodesic}), the equation (\ref{wave}) contains the second derivatives of the particle wavefunction and doesn't gives the class $C^0$ physical solutions.

We want to emphasize that analysis of the initial value problem for the Einstein equations leads to the restriction that admissible coordinate transformations be of class $C^2$ \cite{Lich, Synge}, i.e. all the second order partial derivatives of the metric tensor should exist and be continuous. For the case of discontinuous surfaces, one can impose a weaker condition -- the metric tensors of being of class $C^1$. An admissible coordinate transformation should not change the Riemann tensor, otherwise it will lead to the introduction of the fictitious extra source in the Einstein equations. Indeed, the class $C^0$ coordinate transformations, $x_\nu \to \bar x_\nu$, drop the differentiability of the transformed metric tensor to $C^0$,
\begin{equation}
\bar g_{\mu\nu} = g_{\alpha\beta} \frac {\partial x^\alpha}{\partial \bar x^\mu}\frac {\partial x^\beta}{\partial \bar x^\nu}~,
\end{equation}
even if the initial metric tensor $g_{\alpha\beta}$ was smooth. Then $\bar g_{\mu\nu}$ will introduce the $\delta$-like parts in Riemann tensor, i.e. a singular hyper-surfaces in spacetime. This means that two metric tensors, $g_{\alpha\beta}$ and $\bar g_{\mu\nu}$, are the solutions of different Einstein equations and do not coincide on the surface of discontinuity. To obtain the class $C^0$ (continuous) metric in whole spacetime, one needs to use the Israel-type junction conditions on this singular hyper-surface \cite{Israel}.


Let us as an example consider the most important singular solution of the Einstein equations -- the Schwarzschild metric,
\begin{equation} \label{metric}
ds^2 =  \left(1 - \frac {r_s}{r}\right) dt^2 - \frac {dr^2}{1 - r_s/r} - r^2 d\theta^2 - r^2 sin^2 \theta d\phi^2~,
\end{equation}
where the parameter $r_s = 2MG$ determines the Schwarzschild horizon. To explore the horizon region the standard approach uses the classical geodesic equations (\ref{geodesic}) and introduces so called Regge-Wheeler's tortoise coordinate,
\begin{equation} \label{tortoise}
\bar r = \int \frac {dr}{1-r_s/r} = r + r_s \ln \left( \frac {r}{r_s} - 1\right)~.
\end{equation}
Then in (\ref{geodesic}) the Schwarzschild horizon "disappears" in various singular coordinates and cannot prevent particles to reach the central singularity \cite{BH-1, BH-2, BH-3}. However, the coordinate (\ref{tortoise}), which is the function of the type (\ref{f'}), and the used singular coordinates, like introduced by Kruskal-Szekeres, Eddington-Finkelstein, Lema\^{\i}tre, or Gullstrand-Painlev\'{e}, give $\delta$-functions in the second derivatives, since they contain one of the factors
\begin{equation}
\sqrt{r_s - r}~, \quad {\rm or} \quad \ln |r_s - r|~.
\end{equation}
This means that transformed metric tensors at $r=r_s$ are not differentiable, i.e. are of unacceptable class $C^0$, not of $C^2$, or $C^1$. Then the Einstein equation for these metrics is altered with fictitious $\delta$-sources at $r=r_s$. So, while the singular coordinate transformations (which are necessary to hide the horizon singularity) do not cause problems on the level of the geodesic equations (\ref{geodesic}), they lead to the appearance of $\delta$-functions at $r=r_s$ in the second order differential equations. For instance, the second derivatives of the metric tensor (i.e. the Riemann tensor) enters the equations of motion of a system of particles in the quadruple approximation \cite{Dix},
\begin{equation} \label{quad}
\frac{Dp^\mu}{ds}= F^\mu = - \frac 12 R^\mu{}_{\nu \alpha\beta}u^\nu S^{\alpha \beta} - \frac 16 J^{\alpha\beta\gamma\delta}D^\mu R_{\alpha\beta\gamma\delta}~,
\end{equation}
where $J^{\alpha\beta\gamma\delta}$ is the quadruple moment of the source, $S^{\alpha \beta}$ is the spin tensor and $u^\nu$ is the 4-velocity. Therefore, the force, $F^\mu$, diverges at the Schwarzschild horizon, since the three from six non-zero independent components of the mixed Riemann tensor,
\begin{equation} \label{R=}
R^t{}_{rrt} = 2R^\theta{}_{r\theta r} = 2R^\phi{}_{r\phi r} = \frac {r_s}{r^2(r_s - r)}~,
\end{equation}
blow up at $r = r_s$. At the same time it is known that the Kretschmann invariant for the Schwarzschild metric (\ref{metric}),
\begin{equation} \label{Kretschmann}
R^{\alpha\beta\gamma\delta}R_{\alpha\beta\gamma\delta} = \frac {12 r_s^2}{r^6} ~,
\end{equation}
is seems to be regular at $r = r_s$. However, the expression (\ref{Kretschmann}) is obtained from the assumption of the type $0/0 = 1$  at $r = r_s$. The same is true for the determinant of Schwarzschild's metric tensor, where the product of its components, $g_{tt}\cdot g_{rr}$, is ill-defined at $r = r_s$. In general $g_{tt}$ and $g_{rr}$ are independent functions and the cancelation of their zeros is accidental, since follows from the validity of the vacuum Einstein equations. However, exact spherical symmetry and true vacuums are rarely, if ever, observed.

In our opinion, the conclusion of absence of physical singularities in the points where all geometrical invariants of the Riemann spacetime are regular, is mathematically correct only for at least class $C^2$ metric tensors. For the lower classes, in addition, the behaviors of quantum particles should be considered, since geometry does not exist separately from matter. For instance, in the strong gravitational field close to the Schwarzschild horizon, one can use solutions of the exact second order equation (\ref{wave}), and not only of the classical geodesic equations (\ref{geodesic}). In our previous papers we have explored the equation (\ref{wave}) in Schwarzschild's coordinates, without performing singular coordinate transformations \cite{Gog-Pan, Gog, Gog-Mod}. Using physically boundary conditions at the Schwarzschild horizon, we have found the real-valued exponentially time-dependent solutions (with the complex phases). So quantum particles probably do not enter the horizon around of a compact object, but are absorbed and some are reflected by it. Similar solutions with the complex phase was obtained also in \cite{Tunneling}, were it was nevertheless assumed that classical geodesics are extendable across the horizon. But in this paper the horizon point was removed by the introduction of the infinitesimal integration contours around the propagators pole.


To conclude, in this short article we note that the Regge-Wheeler's tortoise radius and the singular coordinate transformations of metric tensor, used to demonstrate continuity of geodesics at Schwarzschild horizon, are class $C^0$ functions. At the same time, the standard causality theory requires that the metric tensor to be at least $C^1$ function, i.e. whose derivative exists and is continuous. For the case when components of the transformed metric tensor is class $C^0$, its first derivatives are discontinues and the second derivatives lead to the appearance of the fictitious $\delta$-like sources in the Einstein equations. So, while the singular coordinate transformations (which are necessary to hide the horizon singularities) do not cause problems on the level of the classical geodesic equations (which contain the first derivatives of the coordinate functions), they lead to the appearance of $\delta$-functions in the equations of quantum particles (which contain the second derivatives of wavefunctions). The consequence of this observation is that the minimal radius of any isolated body is its Schwarzschild radius and quantum particles probably do not enter the black hole horizon (but are absorbed and some are reflected by it), what potentially can solve some longstand problems. Note that the generally accepted assumption that quantum particles can freely fell through the event horizon contradicts a unitary quantum theory and leads to the another problem -- the black hole information paradox (see the recent review \cite{Mar}).



\begin{thebibliography}{99}

\bibitem{Haw-Pen} S. Hawking and R. Penrose,
                 {\it The Nature of Space and Time}
                 (Princeton University Press, Princeton 1996).

\bibitem{San-Gar} J.M.M. Senovilla and D. Garfinkle,
                "The 1965 Penrose singularity theorem",
                Class. Quant. Grav. {\bf 32} (2015) 124008, arXiv: 1410.5226 [gr-qc].

\bibitem{Haw-Ell} S.W. Hawking and G.F.R. Ellis,
                 {\it The Large Scale Structure of Space-Time}
                 (Cambridge University Press, Cambridge 1973).

\bibitem{Man-San} E. Minguzzi and M. Sanchez,
                 {\it The Causal Hierarchy of Spacetimes in Recent Developments in Pseudo-Riemannian Geometry}, ESI Lect. Math. Phys.
                 (Eur. Math. Soc. Publ. House, Z\"{u}rich 2008).

\bibitem{Clarke} C.J.S. Clarke,
                {\it The Analysis of Spacetime Singularities}, Cambridge Lect. Notes Phys. 1
                (Cambridge University Press, Cambridge 1993).

\bibitem{Gar-San} A. Garc\'{i}a-Parrado and J.M.M. Senovilla,
                 "Causal structures and causal boundaries",
                 Class. Quant. Grav. {\bf 22} (2005) R1, arXiv: gr-qc/0501069.

\bibitem{Chru} P.T. Chru\'{s}ciel,
              "Elements of causality theory",
              arXiv: 1110.6706 [gr-qc].

\bibitem{Sor-Woo} R.D. Sorkin and E. Woolgar,
                 "A causal order for spacetimes with $C^0$ Lorentzian metrics: Proof of compactness of the space of causal curves",
                 Class. Quant. Grav. {\bf 13} (1996) 1971.

\bibitem{Opp-Sny} J.R. Oppenheimer and H. Snyder,
                 "On continued gravitational contraction",
                 Phys. Rev. {\bf 56} (1939) 455.

\bibitem{Lich} A. Lichnerowicz,
              {\it Th\'{e}ories Relativistes de la Gravitation et de l' \'{E}lectromagn\'{e}tisme. Relativit\'{e} G\'{e}n\'{e}rale et Th\'{e}ories Unitaires}
              (Masson, Paris 1955).

\bibitem{Synge} J.L. Synge,
               {\it Relativity: The General Theory}
               (North-Holland Publishing Company, Amsterdam 1960).

\bibitem{Sano} J.M.M. Senovilla,
              "Singularity theorems and their consequences",
              Gen. Rel. Grav. {\bf 30} (1998) 701, arXiv: 1801.04912 [gr-qc].

\bibitem{Gog-Pan} M. Gogberashvili and L. Pantskhava,
                 "Black hole information problem and wave bursts",
                 Int. J. Theor. Phys. {\bf 57} (2018) 1763, arXiv: 1608.04595 [physics.gen-ph].

\bibitem{Gog} M. Gogberashvili,
             "Can quantum particles cross a horizon?",
             arXiv: 1712.02637 [gr-qc].

\bibitem{Gog-Mod} M. Gogberashvili and B. Modrekiladze,
                 "Gravitational field of a spherical perfect fluid",
                 arXiv: 1805.03505 [physics.gen-ph].

\bibitem{Gold} H. Goldstein,
              {\it Classical Mechanics}
              (Addison-Wesley, New York 1950).

\bibitem{Motz} L. Motz and A. Selzer,
              "Quantum mechanics and the relativistic Hamilton-Jacobi equation",
              Phys. Rev. {\bf 133} (1964) B1622.

\bibitem{Israel} W. Israel,
                "Singular hypersurfaces and thin shells in general relativity",
                Nuovo Cim. {\bf B 44} (1966) 1.

\bibitem{BH-1} S. Chandrasekhar,
              {\it The Mathematical Theory of Black Holes}
              (Clarendon, New York 1983).

\bibitem{BH-2} S. Carroll,
              {\it Spacetime and Geometry: An Introduction to General Relativity}
              (Addison-Wesley, San Francisco 2004).

\bibitem{BH-3} E. Poisson,
              {\it A Relativist's Toolkit: The Mathematics of Black-Hole Mechanics}
              (Cambridge University Press, Cambridge 2004).

\bibitem{Dix} W.G. Dixon,
             "The definition of multipole moments for extended bodies",
             Gen. Rel. Grav. {\bf 4} (1973) 199.

\bibitem{Tunneling} K. Srinivasan and T. Padmanabhan,
                   "Particle production and complex path analysis",
                   Phys. Rev. {\bf D 60} (1999) 024007, arXiv: gr-qc/9812028.

\bibitem{Mar} D. Marolf,
             "The black hole information problem: past, present, and future",
             Rep. Prog. Phys. {\bf 80} (2017) 092001, arXiv: 1703.02143 [gr-qc].

\end{thebibliography}
\end{document}